\documentclass[12pt]{article}
\usepackage{pictex,latexsym}
\def\putcircbar at #1 #2 with fuzz #3 {%
  \put {\large $\circ$} at {#1} {#2} 
  \dimen0=\Ydistance{#3}
  \put{\vbox{\hsize=\crossbarlength
    \hrule height \linethickness
    \vskip -.5\linethickness
    \centerline{\vrule width \linethickness height 2\dimen0}
    \nointerlineskip
    \vskip -.5\linethickness
    \hrule height \linethickness}} at {#1} {#2} }
   
\newdimen\xposition
\newdimen\yposition
\newdimen\dyposition
\newdimen\crossbarlength

\def\putdiamondbar at #1 #2 with fuzz #3 {%
  \xposition=\Xdistance{#1}
  \yposition=\Ydistance{#2}
  \dyposition=\Ydistance{#3}

\setdimensionmode
\put {\large $\diamond$} at {\xposition} {\yposition}

\dimen0 = \yposition
  \advance \dimen0 by -\dyposition
\dimen2 = \yposition
  \advance \dimen2  by \dyposition
\putrule from {\xposition} {\dimen0}
  to {\xposition} {\dimen2}

\dimen4 = \xposition
  \advance \dimen4 by -.5\crossbarlength
\dimen6 = \xposition
  \advance \dimen6 by  .5\crossbarlength
\putrule from {\dimen4} {\dimen0} to {\dimen6} {\dimen0}
\putrule from {\dimen4} {\dimen2} to {\dimen6} {\dimen2}
\setcoordinatemode}

\newdimen\xposition
\newdimen\yposition
\newdimen\dyposition
\newdimen\crossbarlength

\def\putbigtriangledownbar at #1 #2 with fuzz #3 {%
  \xposition=\Xdistance{#1}
  \yposition=\Ydistance{#2}
  \dyposition=\Ydistance{#3}

\setdimensionmode
\put {$\bigtriangledown$} at {\xposition} {\yposition}

\dimen0 = \yposition
  \advance \dimen0 by -\dyposition
\dimen2 = \yposition
  \advance \dimen2  by \dyposition
\putrule from {\xposition} {\dimen0}
  to {\xposition} {\dimen2}

\dimen4 = \xposition
  \advance \dimen4 by -.5\crossbarlength
\dimen6 = \xposition
  \advance \dimen6 by  .5\crossbarlength
\putrule from {\dimen4} {\dimen0} to {\dimen6} {\dimen0}
\putrule from {\dimen4} {\dimen2} to {\dimen6} {\dimen2}
\setcoordinatemode}
                                           
\newdimen\xposition
\newdimen\yposition
\newdimen\dyposition
\newdimen\crossbarlength

\def\puttrianglebar at #1 #2 with fuzz #3 {%
  \xposition=\Xdistance{#1}
  \yposition=\Ydistance{#2}
  \dyposition=\Ydistance{#3}

\setdimensionmode
\put {$\triangle$} at {\xposition} {\yposition}

\dimen0 = \yposition
  \advance \dimen0 by -\dyposition
\dimen2 = \yposition
  \advance \dimen2  by \dyposition
\putrule from {\xposition} {\dimen0}
  to {\xposition} {\dimen2}

\dimen4 = \xposition
  \advance \dimen4 by -.5\crossbarlength
\dimen6 = \xposition
  \advance \dimen6 by  .5\crossbarlength
\putrule from {\dimen4} {\dimen0} to {\dimen6} {\dimen0}
\putrule from {\dimen4} {\dimen2} to {\dimen6} {\dimen2}
\setcoordinatemode}
                                          
\newdimen\xposition
\newdimen\yposition
\newdimen\dyposition
\newdimen\crossbarlength

\def\puttrianglerightbar at #1 #2 with fuzz #3 {%
  \xposition=\Xdistance{#1}
  \yposition=\Ydistance{#2}
  \dyposition=\Ydistance{#3}

\setdimensionmode
\put {\large $\triangleright$} at {\xposition} {\yposition}

\dimen0 = \yposition
  \advance \dimen0 by -\dyposition
\dimen2 = \yposition
  \advance \dimen2  by \dyposition
\putrule from {\xposition} {\dimen0}
  to {\xposition} {\dimen2}

\dimen4 = \xposition
  \advance \dimen4 by -.5\crossbarlength
\dimen6 = \xposition
  \advance \dimen6 by  .5\crossbarlength
\putrule from {\dimen4} {\dimen0} to {\dimen6} {\dimen0}
\putrule from {\dimen4} {\dimen2} to {\dimen6} {\dimen2}
\setcoordinatemode}
                        
\newdimen\xposition
\newdimen\yposition
\newdimen\dyposition

\def\puttriangleleftbar at #1 #2 with fuzz #3 {%
  \xposition=\Xdistance{#1}
  \yposition=\Ydistance{#2}
  \dyposition=\Ydistance{#3}

\setdimensionmode
\put {\large $\triangleleft$} at {\xposition} {\yposition}

\dimen0 = \yposition
  \advance \dimen0 by -\dyposition
\dimen2 = \yposition
  \advance \dimen2  by \dyposition
\putrule from {\xposition} {\dimen0}
  to {\xposition} {\dimen2}

\dimen4 = \xposition
  \advance \dimen4 by -.5\crossbarlength
\dimen6 = \xposition
  \advance \dimen6 by  .5\crossbarlength
\putrule from {\dimen4} {\dimen0} to {\dimen6} {\dimen0}
\putrule from {\dimen4} {\dimen2} to {\dimen6} {\dimen2}
\setcoordinatemode}

\newcommand{\AmS}{{\protect\the\textfont2
  A\kern-.1667em\lower.5ex\hbox{M}\kern-.125emS}}

\begin{document}
\begin{titlepage}
\pagestyle{empty}
\begin{center}
\today \hfill IPNO/TH 95-48\\
\hfill hep-lat/9805010\\

\vskip .3in

{\Large
Noncompact, Gauge-Invariant Simulations of $U(1)$, $SU(2)$, and $SU(3)$
\footnote{Research supported by the U.~S. Department of Energy
under contracts DE-FG04-84ER40166 and 
DE-FG03-92ER40732/A004 (task B).}}

\vskip .3in

Kevin Cahill\\

\vskip .10in
\baselineskip=10pt
{\em Department of Physics and Astronomy,
        University of New Mexico\\ 
        Albuquerque, New Mexico 87131-1156, U.~S.~A.
\footnote{Permanent address, e-mail: kevin@cahill.phys.unm.edu}
\\
\vskip .08in
Division de Physique Th\'eorique,
\footnote{Unit\'e de Recherche des Universit\'es Paris XI
et Paris VI associ\'ee au CNRS}      
Institut de Physique Nucl\'eaire\\
91406 Orsay Cedex, France.}

\vskip .20in

Gary Herling\\

\vskip .10in

{\em Center for Advanced Studies\\
Department of Physics and Astronomy, University of New Mexico\\
Albuquerque, New Mexico 87131-1156, U.~S.~A.
\footnote{E-mail: herling @ unm.edu}}
\baselineskip=26pt
\vskip 0.4in

\begin{abstract}
We have applied a new noncompact, gauge-invariant, 
Monte Carlo method to simulate
the $U(1)$, $SU(2)$, and $SU(3)$ gauge theories
on $8^4$ and $12^4$ lattices.
For $U(1)$ the Creutz ratios of the Wilson loops
agree with the exact results 
for $\beta \ge 0.5$
after a renormalization of the charge.
The $SU(2)$ and $SU(3)$ Creutz ratios
robustly display quark confinement at $\beta = 0.5$
and $\beta = 2$, respectively.
At much weaker coupling, the $SU(2)$ and $SU(3)$ Creutz ratios
agree with perturbation theory after a renormalization
of the coupling constant.
For $SU(3)$ without quarks,
our lattice QCD parameter is $ \Lambda_L = 130 \pm 18$ MeV\null.
\end{abstract}
\baselineskip=16pt
\end{center}
\end{titlepage}
\newpage
\renewcommand{\thepage}{\arabic{page}}
\setcounter{page}{1}
\goodbreak

\section{Introduction}

In compact lattice gauge theory,
gauge fields are represented by group elements
rather than by fields,
and the action is a {\it periodic\/} function
of a gauge-invariant lattice field strength.
The periodicity of the action entails spurious vacua.  
The principal advantage of noncompact actions,
in which gauge fields are represented by fields,
is that they avoid multiple vacua.
\par
Palumbo, Polikarpov, and Veselov~\cite{Palumbo}
carried out 
the first gauge-invariant noncompact simulations.
They saw a confinement signal.
Their action contains five terms,
constructed from two invariants,
and involves (noncompact) auxiliary fields
and an adjustable parameter.
\par
The present paper describes a test
of a new way~\cite{CahGary}
of performing gauge-invariant
noncompact simulations.
Our action, which is
similar to one term
of Palumbo's action,
is exactly invariant under compact
gauge transformations,
is a natural discretization
of the classical Yang-Mills action,
and reduces to Wilson's action
when the gauge fields are compactified.
In our version of Palumbo's method,
there are fewer auxiliary fields, 
and they are compact group elements related to 
gauge transformations. 
\par
We have used this method
to simulate the $U(1)$, $SU(2)$, and $SU(3)$ gauge theories
on $8^4$ and $12^4$ lattices.
For $U(1)$ our Creutz ratios of Wilson loops
agree with the exact ratios of the free continuum theory
after a renormalization of the charge
when the inverse coupling $ \beta $
exceeds $ 0.5 $\null.
Our $SU(2)$ and $SU(3)$ Creutz ratios
clearly show quark confinement at $\beta = 0.5$
and $\beta = 2$, respectively.
At much weaker coupling, our $SU(2)$ and $SU(3)$ Creutz ratios
approximate those of continuum perturbation theory 
when the coupling constant
is suitably renormalized.
For $SU(3)$ there is a scaling 
in the transition region $ 2.2 \le \beta \le 3 $
with a lattice QCD parameter 
$ \Lambda_L \approx 130 \pm 18 $ MeV\null,
which is to be compared with the continuum value 
$ \Lambda_{\rm \overline{\rm MS}}^{(0)} \approx 210 $ MeV\null. 
                                            

\section{The Method}

For massless fermions, the continuum action density is
$\bar \psi i \gamma_\mu \partial_\mu \psi.$
A suitable discretization of this quantity is
$i \bar \psi (n) \gamma_\mu
[ \psi(n + e_\mu) - \psi(n)]/a$
in which $n$ is a four-vector
of integers representing an arbitrary vertex
of the lattice, $e_\mu$ is a unit vector
in the $\mu$th direction, and
$a$ is the lattice spacing.
The product of Fermi fields at the same
point is gauge invariant as it stands.
The other product of Fermi fields
becomes gauge invariant
if we insert a matrix $A_\mu(n)$ of gauge fields
\begin{equation}
\bar \psi (n) \gamma_\mu
\left[ 1 + i g a A_\mu(n) \right] \psi(n + e_\mu)
\end{equation}
that transforms appropriately.
Under a gauge transformation
represented by the group elements
$U(n)$ and $U(n + e_\mu)$,
the required response is
\begin{equation}
1 + i a g A'_\mu(n) =
U(n) [ 1 + i a g A_\mu(n) ] U^{-1}(n + e_\mu)
\label{1+A'}
\end{equation}
or equivalently
\begin{equation}
A'_\mu(n) = U(n) A_\mu(n) U^{-1}(n + e_\mu)
+ {i \over a g} U(n)
\left[ U^{-1}(n) - U^{-1}(n + e_\mu) \right].
\label{A'}
\end{equation}
Under this gauge transformation, 
the lattice field strength 
\begin{eqnarray}
F_{\mu\nu}(n) & = &
{ 1 \over a } \left[ A_\mu(n+e_\nu) - A_\mu(n) \right]
- { 1 \over a } \left[ A_\nu(n+e_\mu) - A_\nu(n) \right]
\nonumber \\
& & \mbox{} + i g  \left[ A_\nu(n) A_\mu(n+e_\nu) 
- A_\mu(n) A_\nu(n+e_\mu) \right],
\label{F}
\end{eqnarray}
which reduces to the continuum Yang-Mills
field strength in the limit $a \to 0$,
transforms as
\begin{equation}
F'_{\mu\nu}(n) = U(n) F_{\mu\nu}(n) U^{-1}(n + e_\mu + e_\nu).
\label{F'}
\end{equation}
The field strength $F_{\mu\nu}(n)$
is antisymmetric
in the indices $\mu$ and $\nu$, but it is not
hermitian.
To make a positive plaquette action density,
we use the Hilbert-Schmidt norm of $F_{\mu\nu}(n)$
\begin{equation}
S = {1 \over 4 k} {\rm Tr} [F^\dagger_{\mu\nu}(n) F_{\mu\nu}(n)],
\label{S}
\end{equation}
in which the generators $T_a$
of the gauge group are normalized as
${\rm Tr} ( T_a T_b ) = k \delta_{ab}$\null.
Because $F_{\mu\nu}(n)$ transforms
covariantly (\ref{F'}), this action density
is exactly invariant under the
noncompact gauge transformation (\ref{A'}).
\par
The gauge transformation (\ref{A'})
with group element $U(n) = \exp(-i a g \omega^a T_a )$
typically maps the matrix of gauge fields
$A_\mu(n) = T_a A^a_\mu(n)$
outside the Lie algebra,
apart from terms of lowest (zeroth) order in the
lattice spacing $a$\null.
We accept this larger space of matrices.
We use the action (\ref{S})
in which the field strength (\ref{F})
is defined in terms of gauge-field
matrices $ A_\mu(n) $ that are
the images under arbitrary gauge transformations
of matrices $ A^0_\mu(n) $ of gauge fields
defined in the usual way,
$ A^0_\mu(n) \equiv T_a A^{a,0}_\mu(n) $
where the fields $ A^{a,0}_\mu(n) $ are real:
\begin{equation}
1 + i a g A_\mu(n)
= V_\mu(n) [ 1 + i a g A^0_\mu(n) ] W_\mu^{-1}(n+e_\mu).
\label{newA'}
\end{equation}
These gauge transformations 
are applied separately and link-wise
to the gauge-field matrices $ A^0_\mu(n) $. 
In general the group elements $V_\mu(n)$ and $ W_\mu(n+e_\mu) $
associated with the gauge field $A_\mu(n)$ 
are unrelated to those
associated with the neighboring
gauge fields $ A_\mu(n+e_\nu) $,
$ A_\nu(n) $, and $ A_\nu(n+e_\mu) $.
But when the matrices $ V_\mu $ and $ W_\mu $ are equal
and independent of direction
\begin{equation}
 V_\mu (n) = W_\mu (n) = U(n) 
\label{gauge condition}
\end{equation}
for every vertex $ n $, then they constitute
a gauge transformation. 
\par
Actually there is only one independent group element
associated with each link.
For by writing the equation (\ref{newA'})
in the form
\begin{eqnarray}
1 + i a g A_\mu(n) 
& = & V_\mu(n)
[ 1 + i a g A^0_\mu(n) ] V^{-1}_\mu(n) \nonumber \\
& & \mbox{}
\times V_\mu(n) W_\mu^{-1}(n+e_\mu),
\label{newVWA'}
\end{eqnarray}
we see that the departure of the gauge-field matrix
$ A_\mu(n) $ from the Lie algebra is entirely
due to the last matrix product $ V_\mu(n) W_\mu^{-1}(n+e_\mu) $\null.
Thus there are as many auxiliary fields 
in this method as there are generators
of the gauge group.
For $ SU(N) $,  the product $ VW^{-1} $ has
$ N^2 -1 $ generators, so our method
involves $ N^2 -1 $ auxiliary fields
in this case.
We have not tried to parameterize these products 
and have instead accepted
whatever matrices the link-wise gauge transformation~(\ref{newA'})
generated.
\par
Palumbo, in his version of this method~\cite{Palumbo},
proposed a parameterization
for the gauge-field matrices $ A_\mu $
that is slightly more general than is necessary and 
that leads to two more auxiliary fields than in our version.
For instance, in the case of $ SU(N) $
his procedure involves $ N^2 + 1 $ auxiliary fields
which support a $ U(N) $ gauge invariance.
His action also has a special term 
that may be needed to suppress these extra
auxiliary fields.
\par
The auxiliary fields that describe the group element
$ VW^{-1} $ are the principal defect
of the present method.  
Their presence may be detected by measuring the average value
of the path-ordered products of the
factors $ V_\mu(n)W_\mu^{-1}(n+e_\mu) $ around the plaquettes
of the lattice.
For when the gauge-transformation
condition (\ref{gauge condition})
is satisfied, the product around each plaquette 
\begin{eqnarray}
P_{\mu\nu}(n) & = & V_\mu(n) W_\mu^{-1}(n+e_\mu) 
V_\nu(n+e_\mu) W_\nu^{-1}(n+e_\mu + e_\nu )
\nonumber \\
& & \mbox{}
\times
[ V_\mu(n+ e_\nu ) W_\mu^{-1}(n+e_\mu+ e_\nu ) ]^\dagger
[ V_\nu(n) W_\nu^{-1}(n + e_\nu ) ]^\dagger
\label{VWplaquette}
\end{eqnarray}
is unity.
\par
To estimate the effects of the auxiliary fields,
we have measured the mean values of these products
\begin{equation}
\langle P \rangle = { 1 \over 6 L^4 } \sum_{n,\mu\nu}
\langle P_{\mu\nu}(n) \rangle
\label{P}
\end{equation}
at various values of $ \beta $ in our $ SU(3) $ simulations.
We found that the auxiliary fields do reduce to
gauge transformations in the continuum limit,
$ \beta \to \infty $, but slowly:
at $ \beta = 2 $, $ \langle P \rangle = 0.18 $;
at $ \beta = 3$, $ \langle P \rangle = 0.30 $;
at $ \beta = 100$, $ \langle P \rangle = 0.71 $;
and at $ \beta = 1000$, $ \langle P \rangle = 0.87 $.
\par
The quantity
$ 1 + iga A_\mu(n) $
is not an element $ L_\mu(n) $
of the gauge group,
except when the real gauge fields
$ A_\mu^{a,0}(n) $ all vanish.
But if one compactified
the fields by requiring $ 1 + iga A_\mu(n) $
to be an element of the gauge group,
then the matrix $ A_\mu(n) $
of gauge fields would be
related to the link $ L_\mu(n) $ by
$A_\mu(n) =  ( L_\mu(n) - 1 )/( iga )$,
and the action (\ref{S})
defined in terms of the field strength
(\ref{F}) would be, {\it mirabile dictu\/},
Wilson's action:
\begin{equation}
S =
{k - \Re \, {\rm Tr} L_\mu(n) L_\nu(n+e_\mu)
L^\dagger_\mu(n+e_\nu) L^\dagger_\nu(n)
\over 2 a^4 g^2k }.
\nonumber
\label{W}
\end{equation}
                       
\section{Results}

We have tested this method by applying it
to the $U(1)$, $SU(2)$, and $SU(3)$ gauge
theories on $8^4$ and $12^4$ lattices.
We used a multi-hit Metropolis algorithm
with a flat measure for the real gauge
fields $ A^{a,0}_\mu(n) $ and the relevant
Haar measure for the group elements
$ V_\mu(n) $ and $ W_\mu(n) $.
In most of our initial configurations,
the unitary matrices $V$ and $W$ 
and the hermitian gauge fields $A^0_\mu$
were randomized.
For thermalization
we allowed 50,000 sweeps for $U(1)$,
10,000 for $SU(2)$, and 100,000
for $SU(3)$.
We allowed at least 20 sweeps between measurements.
Our Wilson loops are ensemble averages
of ordered products of the binomials
$1 + i a g A_\mu(n)$ 
rather than of the exponentials
$\exp[i a g A_\mu(n)]$ around the loop. 
\par
For $U(1)$ and for $\beta = 1/g^2 \ge 0.5$,
our measured Creutz ratios~\cite{Creu80b}
of Wilson loops
agree with the exact ones
of the free continuum theory
apart from finite-size effects
and a renormalization of the charge.
For instance at $\beta = 0.5$
on the $12^4$ lattice, we found
$\chi(2,2) =  0.147(1)$,
$\chi(2,3) =  0.103(1)$,
$\chi(2,4) =  0.090(1)$,
$\chi(3,3) =  0.049(1)$,
$\chi(3,4) =  0.034(1)$, and
$\chi(4,4) =  0.020(2)$.
At the value $\beta_r = 0.45 $
of the renormalized charge,
the exact Creutz ratios of the continuum theory
are:
$\chi(2,2) =  0.147$,
$\chi(2,3) =  0.102$,
$\chi(2,4) =  0.093$,
$\chi(3,3) =  0.048$,
$\chi(3,4) =  0.037$, and
$\chi(4,4) =  0.024$.
The agreement is excellent
apart from the two largest loops
which are too small by 8\%
and by 17\%, respectively,
largely due to finite-size effects.
\par
At stronger coupling,
the extra terms
\begin{equation}
i g  \left[ A_\nu(n) A_\mu(n+e_\nu)
- A_\mu(n) A_\nu(n+e_\mu) \right]
\label{extra terms}
\end{equation}
in the lattice field
strength $F_{\mu\nu}(n)$
eventually do produce a
confinement signal.
For example, at $\beta = 0.375 $,
our measured Creutz ratios
on the $12^4$ lattice are:
$\chi(2,2) =  0.906(5)$,
$\chi(2,3) =  0.909(21)$,
$\chi(2,4) =  0.85(10)$,
$\chi(3,3) =  0.62(24)$, and
$\chi(3,4) =  0.6(16)$.
\par
For $SU(2)$
on the $8^4$ lattice at $\beta = 4/g^2 = 0.5$,
we found 
$\chi(2,2) =  0.835(3)$,
$\chi(2,3) =  0.852(12)$,
$\chi(2,4) =  0.865(60)$, and
$\chi(3,3) =  0.94(23)$
which within the limited statistics
clearly exhibit confinement.
\par
To test this method at weaker coupling,
we compared our Creutz ratios with
those of continuum perturbation theory.
The tree-level perturbative formula
for $ SU(n) $ may be expressed~\cite{Cahill}  
in terms of the function $ u(i,j) $ 
\begin{equation}
u(i,j) = {i \over j} \arctan{ i \over j }
+ { j \over i } \arctan{ j \over i }
- \log\left( { 1 \over i^2 } + { 1 \over j^2 } \right)
\label{u}
\end{equation}
as
\begin{equation}
\chi(i,j) = { n^2 - 1 \over 2 \pi^2 \beta }
\left[ - u(i,j) - u(i-1,j-1) 
+ u(i,j-1) + u(i-1,j) \right].
\label{tree}
\end{equation}
\par
For $ SU(2) $ at $\beta = 1$, our six Creutz ratios
$\chi(2,2) =  0.1087(5)$,
$\chi(2,3) =  0.0783(5)$,
$\chi(2,4) =  0.0696(5)$,
$\chi(3,3) =  0.0414(7)$,
$\chi(3,4) =  0.0299(8)$, and
$\chi(4,4) =  0.0189(10)$
are close to the $ \chi $'s of the perturbative formula (\ref{tree}) 
at the renormalized value of $\beta_r = 1.75$, to wit: 
$\chi(2,2) =  0.1133$,
$\chi(2,3) =  0.0786$,
$\chi(2,4) =  0.0721$,
$\chi(3,3) =  0.0374$,
$\chi(3,4) =  0.0286$, and
$\chi(4,4) =  0.0187$.
\par
For $SU(3)$ at $\beta = 6/g^2 = 2$
on the $12^4$ lattice,
we found in ten independent runs 
$\chi(2,2) =  0.838(1)$,
$\chi(2,3) =  0.826(3)$,
$\chi(2,4) =  0.828(13)$,
$\chi(3,3) =  0.793(42)$,
$\chi(3,4) =  0.47(25)$, and
$\chi(4,4) =  1.2(86)$.
Within the statistics,
these results robustly exhibit confinement.
\par
At much weaker coupling, our ratios
approximate those of the tree-level formula (\ref{tree})
of continuum perturbation theory
apart from finite-size effects and
after a renormalization of the coupling constant.
At $ \beta = 4 $ on the $ 8^4 $ lattice, for instance,
we found in one run 
$\chi(2,2) =  0.0878(1)$,
$\chi(2,3) =  0.0622(1)$,
$\chi(2,4) =  0.0554(2)$,
$\chi(3,3) =  0.0319(2)$,
$\chi(3,4) =  0.0228(6)$, and
$\chi(4,4) =  0.0132(1)$;
whereas the tree-level formula (\ref{tree}) 
at the renormalized value of $ \beta_r = 6.03 $ gives
$\chi(2,2) =  0.0878$,
$\chi(2,3) =  0.0609$,
$\chi(2,4) =  0.0559$,
$\chi(3,3) =  0.0290$,
$\chi(3,4) =  0.0222$, and
$\chi(4,4) =  0.0145$\null.
Similarly at $ \beta = 100 $ we found 
in one run
$\chi(2,2) =  0.00477(1)$,
$\chi(2,3) =  0.00325(1)$,
$\chi(2,4) =  0.00289(1)$,
$\chi(3,3) =  0.00150(1)$,
$\chi(3,4) =  0.00104(2)$, and
$\chi(4,4) =  0.00053(3)$;
whereas the perturbative formula (\ref{tree}) 
at $ \beta_r = 111 $ gives
$\chi(2,2) =  0.00476$,
$\chi(2,3) =  0.00331$,
$\chi(2,4) =  0.00303$,
$\chi(3,3) =  0.00157$,
$\chi(3,4) =  0.00120$, and
$\chi(4,4) =  0.00078$\null.
The better agreement for the smaller loops
is a finite-size effect.

\section{Scaling}
\par
We used an $8^4$ lattice to 
study the scaling of the lattice
spacing $a$ with the coupling constant $g$
for $SU(3)$.  The two-loop result
for the dependence of the string tension
$\sigma a^2$ upon the inverse coupling
$ \beta $ is
\begin{equation}
\sigma a^2 \approx
{ \sigma \over \Lambda_L^2 } 
\exp \left[ - { 8 \pi^2 \beta \over 33 }
+ {102 \over 121} \log\left( { 8 \pi^2 \beta \over 33 } \right)
\right].
\label{scaling}
\end{equation}
\par
We expect this scaling formula to hold
in a transition region where perturbation theory
is still valid and where the quark-antiquark
static potential $ V(r) $ is a linear combination
of a confining linear potential and a Coulomb potential. 
At one end of this region,
$ V(r) $ is mostly linear;
at the other end it, is mostly Coulomb.
We found a transition region
between $ \beta = 2.2 $ and $ \beta = 3 $\null.
In this region,
we used the interpolation
\begin{equation}
\chi(i,j) = {1 \over 3 }
\left[ ( 4 - \beta ) \sigma a^2
+ ( \beta -1 ) \chi_0(i,j) \right]
\label{interpol}
\end{equation}
in which the string tension $ \sigma a^2 $ is 
given by the scaling formula (\ref{scaling})
and the perturbative Creutz ratio $ \chi_0(i,j) $
is given by the tree-level formula (\ref{tree}).
\par
In the figure we have plotted the Creutz ratios
$ \chi(i,j) $ that we measured on an $ 8^4 $ lattice
for $ 1.5 \le \beta \le 3 $\null.
We also have plotted the interpolative formula (\ref{interpol})
for $ 2.2 \le \beta \le 3 $
for various values of $ \sigma/\Lambda^2 $ 
from 8 to 14 as solid curves.
Our $ 2 \times 2 $ Creutz ratios
fit the interpolation (\ref{interpol})
in the transition region $ 2.2 \le \beta \le 3 $
with $ \sigma/\Lambda_L^2 = 14 $\null.
Our ratios $ \chi(2,3) $ and $ \chi(2,4) $ fit it
with $ \sigma/\Lambda_L^2 = 12 $ and 11, respectively.
Our ratios $ \chi(3,3) $ fit it
with $ \sigma/\Lambda_L^2 = 8 $\null.
Altogether our $ \chi $'s fit the
interpolation (\ref{interpol})
for $ 2.2 \le \beta \le 3 $
with $ \sigma/\Lambda_L^2 = 11 \pm 3 $\null.
A string tension 
$\sqrt{\sigma} \approx 420$ MeV
implies that
$ \Lambda_L \approx 130 \pm 18 $ MeV,
which is to be compared with
the continuum value of
$ \Lambda_{\overline{\rm MS}}^{(0)} \approx 210 $ MeV
and with the lattice 
parameter $ \Lambda_{LW} \approx 7.9$ MeV
of Wilson's method.
\par
At stronger coupling where the transition
to a linear potential is complete,
our $\chi(i,j)$'s fit the scaling formula (\ref{scaling})
without any Coulomb term for $ 1.9 < \beta < 2.1 $
if we set $ \sigma/\Lambda_L^2 \approx 25.0 \pm 4 $. 
The corresponding value of $ \Lambda_L $ is
$ \Lambda_L \approx 85 \pm 7 $ MeV.

\smallskip
\section*{Acknowledgments}
\par
We are indebted to
M.~Creutz,
G.~Marsaglia, F.~Palumbo, W.~Press, and K.~Webb
for useful conversations,
to the Department of Energy for financial support
under task B of grant DE-FG03-92ER40732/A004, and to 
B.~Dieterle and the Maui Center for High-Performance
Computing\footnote[1]
{Research sponsored in part by the Phillips Laboratory, Air
     Force Materiel Command, USAF, under cooperative agreement
     F29601-93-2-0001.  The U.S. Government retains a
     nonexclusive copyright to this work.  The views and conclusions
     of this work are those of the authors and
     should not be interpreted as necessarily representing the
     official policies or endorsements, either expressed or
     implied, of Phillips Laboratory or the U.S. Government.}
for computer time.
K.~Cahill would like to thank A.~Comtet and D.~Vautherin
for inviting him to Orsay.
\bibliographystyle{unsrt}

\par
\section*{Figure Caption}
The $SU(3)\/$ Creutz ratios $\chi(i,j)$ and
the scaling predictions for the string tension $\protect \sigma a^2$
are plotted against $\beta$.  
The solid lines for $1.9 < \beta < 2.1$
represent formula (\ref{scaling}) for $\protect \sigma a^2$
with $ \protect \sigma/\Lambda_L^2 = 25 \pm 4 $\null.
The solid curves for $2.2 < \beta < 3.0$
are the interpolation (\ref{interpol}) between the
tree-level formula (\ref{tree}) for the Creutz ratios $ \chi(i,j) $
and the scaling prediction (\ref{scaling})
for $\protect \sigma a^2$ with 
$ \protect \sigma/\Lambda_L^2 = 11 \pm 3 $\null.

\vfill \eject
\onecolumn
\large

\begin{figure} [htb]
\beginpicture
\crossbarlength=5pt

\ticksin
\inboundscheckon
\setcoordinatesystem units <2.80in,2.60in>   
\setplotarea x from 1.4 to 3.1, y from -2.0 to 0.47712
\axis bottom label {$\beta$} ticks
  numbered at 1.5 2.0 2.5 3.0 /
  unlabeled short from 1.6 to 1.9 by 0.1 
  from 2.1 to 2.4 by 0.1 from 2.6 to 2.9 by 0.1 /
\axis left  label {$\chi$}
  ticks logged
  numbered at 0.01 0.1 1.0 3.0 /
  unlabeled short at 2.0 /
  unlabeled short from 2.0 to 2.0 by 1.0
  from 0.2 to 0.9 by 0.1
  from 0.02 to 0.09 by 0.01 /
\axis top label {}
  ticks
  unlabeled at 1.5 2.0 2.5 3.0 /
  unlabeled short from 1.6 to 1.9 by 0.1
  from 2.1 to 2.4 by 0.1 from 2.6 to 2.9 by 0.1 /
\axis right ticks logged
  unlabeled at 0.1 1.0 /
  unlabeled short from 2.0 to 2.0 by 1.0
  unlabeled short from 0.2 to 0.9 by 0.1
  from 0.02 to 0.09 by 0.01 /

\put {$SU(3)$ \quad $8^4$} at 2.5 0.2

\setquadratic
\setsolid
\plot  1.9 -0.097718  2.0 -0.182850  2.1 -0.268899 /
\plot  1.9 0.042461  2.0 -0.042671  2.1 -0.128720 /
\setsolid
\plot 2.2 -0.564748 2.6 -0.771949 3.0 -0.865705 /
\plot 2.2 -0.661884 2.6 -0.895441 3.0 -1.010508 /
\plot 2.2 -0.699583 2.6 -0.933057 3.0 -1.048052 /
\plot  2.2 -0.877825 2.6 -1.152120 3.0 -1.305291 /
\setsolid

\inboundscheckon

\putcircbar at  1.500  0.068635 with fuzz  0.000889
\puttrianglebar at  1.500  0.066000 with fuzz  0.004593
\putdiamondbar at  1.500 -0.252850 with fuzz  0.176371
\putbigtriangledownbar at  1.500  0.101571 with fuzz  0.045712

\putcircbar at  1.600  0.046974 with fuzz  0.000895
\puttrianglebar at  1.600  0.044917 with fuzz  0.004748
\putdiamondbar at  1.600 -0.191106 with fuzz  0.148440
\putbigtriangledownbar at  1.600  0.016430 with fuzz  0.039381

\putcircbar at  1.700  0.025898 with fuzz  0.000668
\puttrianglebar at  1.700  0.027521 with fuzz  0.003164
\putdiamondbar at  1.700  0.176648 with fuzz  0.086632
\putbigtriangledownbar at  1.700  0.045226 with fuzz  0.023665

\putcircbar at  1.800 -0.000986 with fuzz  0.000596
\puttrianglebar at  1.800 -0.002540 with fuzz  0.003039
\putdiamondbar at  1.800 -0.019013 with fuzz  0.056526
\putbigtriangledownbar at  1.800 -0.022986 with fuzz  0.016663

\putcircbar at  1.900 -0.033710 with fuzz  0.000520
\puttrianglebar at  1.900 -0.037287 with fuzz  0.002062
\putdiamondbar at  1.900 -0.031785 with fuzz  0.037028
\putbigtriangledownbar at  1.900 -0.028525 with fuzz  0.012268

\putcircbar at  1.950 -0.054826 with fuzz  0.000595
\puttrianglebar at  1.950 -0.060990 with fuzz  0.002188
\putdiamondbar at  1.950 -0.087482 with fuzz  0.030696
\putbigtriangledownbar at  1.950 -0.062335 with fuzz  0.009190
\puttrianglerightbar at  1.950 -0.152009 with fuzz  0.213403

\putcircbar at  2.000 -0.078260 with fuzz  0.000344
\puttrianglebar at  2.000 -0.083529 with fuzz  0.000869
\putdiamondbar at  2.000 -0.086209 with fuzz  0.011930
\putbigtriangledownbar at  2.000 -0.083520 with fuzz  0.003536
\puttrianglerightbar at  2.000 -0.244795 with fuzz  0.095973

\putcircbar at  2.050 -0.111278 with fuzz  0.000527
\puttrianglebar at  2.050 -0.118801 with fuzz  0.001086
\putdiamondbar at  2.050 -0.135791 with fuzz  0.010435
\putbigtriangledownbar at  2.050 -0.119095 with fuzz  0.003333
\puttrianglerightbar at  2.050 -0.099997 with fuzz  0.064153

\putcircbar at  2.100 -0.160707 with fuzz  0.000464
\puttrianglebar at  2.100 -0.173773 with fuzz  0.000703
\putdiamondbar at  2.100 -0.191438 with fuzz  0.005593
\putbigtriangledownbar at  2.100 -0.178051 with fuzz  0.001594
\puttrianglerightbar at  2.100 -0.231290 with fuzz  0.021762

\putcircbar at  2.150 -0.418639 with fuzz  0.004482
\puttrianglebar at  2.150 -0.484180 with fuzz  0.005664
\putdiamondbar at  2.150 -0.599664 with fuzz  0.008142
\putbigtriangledownbar at  2.150 -0.510715 with fuzz  0.006288
\puttrianglerightbar at  2.150 -0.650035 with fuzz  0.009133
\puttriangleleftbar at  2.150 -0.695546 with fuzz  0.025359

\putcircbar at  2.200 -0.560824 with fuzz  0.000667
\puttrianglebar at  2.200 -0.661491 with fuzz  0.001017
\putdiamondbar at  2.200 -0.847092 with fuzz  0.002147
\putbigtriangledownbar at  2.200 -0.699081 with fuzz  0.001263
\puttrianglerightbar at  2.200 -0.933639 with fuzz  0.003216
\puttriangleleftbar at  2.200 -1.066131 with fuzz  0.007334

\putcircbar at  2.300 -0.644382 with fuzz  0.000345
\puttrianglebar at  2.300 -0.763208 with fuzz  0.000512
\putdiamondbar at  2.300 -0.989603 with fuzz  0.001104
\putbigtriangledownbar at  2.300 -0.808789 with fuzz  0.000630
\puttrianglerightbar at  2.300 -1.103573 with fuzz  0.001741
\puttriangleleftbar at  2.300 -1.286561 with fuzz  0.004628

\putcircbar at  2.400 -0.695841 with fuzz  0.000664
\puttrianglebar at  2.400 -0.820857 with fuzz  0.001039
\putdiamondbar at  2.400 -1.057601 with fuzz  0.002835
\putbigtriangledownbar at  2.400 -0.866909 with fuzz  0.001446
\puttrianglerightbar at  2.400 -1.172853 with fuzz  0.004507
\puttriangleleftbar at  2.400 -1.364710 with fuzz  0.009844

\putcircbar at  2.500 -0.738825 with fuzz  0.000605
\puttrianglebar at  2.500 -0.869488 with fuzz  0.000916
\putdiamondbar at  2.500 -1.121339 with fuzz  0.002662
\putbigtriangledownbar at  2.500 -0.919019 with fuzz  0.001382
\puttrianglerightbar at  2.500 -1.247846 with fuzz  0.004019
\puttriangleleftbar at  2.500 -1.449511 with fuzz  0.009404

\putcircbar at  2.600 -0.777584 with fuzz  0.000699
\puttrianglebar at  2.600 -0.912775 with fuzz  0.001220
\putdiamondbar at  2.600 -1.176549 with fuzz  0.003060
\putbigtriangledownbar at  2.600 -0.962278 with fuzz  0.001449
\puttrianglerightbar at  2.600 -1.308541 with fuzz  0.004890
\puttriangleleftbar at  2.600 -1.538306 with fuzz  0.012895

\putcircbar at  2.700 -0.807254 with fuzz  0.001105
\puttrianglebar at  2.700 -0.944531 with fuzz  0.001548
\putdiamondbar at  2.700 -1.208071 with fuzz  0.002405
\putbigtriangledownbar at  2.700 -0.993339 with fuzz  0.002034
\puttrianglerightbar at  2.700 -1.340910 with fuzz  0.005859
\puttriangleleftbar at  2.700 -1.574302 with fuzz  0.011224

\putcircbar at  2.800 -0.834246 with fuzz  0.001089
\puttrianglebar at  2.800 -0.970910 with fuzz  0.002117
\putdiamondbar at  2.800 -1.230601 with fuzz  0.007057
\putbigtriangledownbar at  2.800 -1.020515 with fuzz  0.001957
\puttrianglerightbar at  2.800 -1.361886 with fuzz  0.009007
\puttriangleleftbar at  2.800 -1.583422 with fuzz  0.013033

\putcircbar at  2.900 -0.861091 with fuzz  0.000837
\puttrianglebar at  2.900 -1.000642 with fuzz  0.001310
\putdiamondbar at  2.900 -1.268458 with fuzz  0.003802
\putbigtriangledownbar at  2.900 -1.050769 with fuzz  0.001388
\puttrianglerightbar at  2.900 -1.402858 with fuzz  0.004672
\puttriangleleftbar at  2.900 -1.611191 with fuzz  0.011385

\putcircbar at  3.000 -0.882830 with fuzz  0.001171
\puttrianglebar at  3.000 -1.023488 with fuzz  0.002116
\putdiamondbar at  3.000 -1.291547 with fuzz  0.005902
\putbigtriangledownbar at  3.000 -1.071911 with fuzz  0.002572
\puttrianglerightbar at  3.000 -1.419731 with fuzz  0.011565
\puttriangleleftbar at  3.000 -1.626795 with fuzz  0.025939

\put {$\circ$  $\chi(2,2)$ } at 1.7 -0.95
\put {{\tiny $\triangle$}  $\chi(2,3)$ } at 1.7 -1.07
\put {{\tiny $\bigtriangledown$}  $\chi(2,4)$ } at 1.7 -1.19
\put {$\diamond$  $\chi(3,3)$ } at 1.7 -1.31
\put {$\triangleright$  $\chi(3,4)$ } at 1.7 -1.43
\put {$\triangleleft$  $\chi(4,4)$ } at 1.7 -1.55

%

\endpicture
\end{figure}

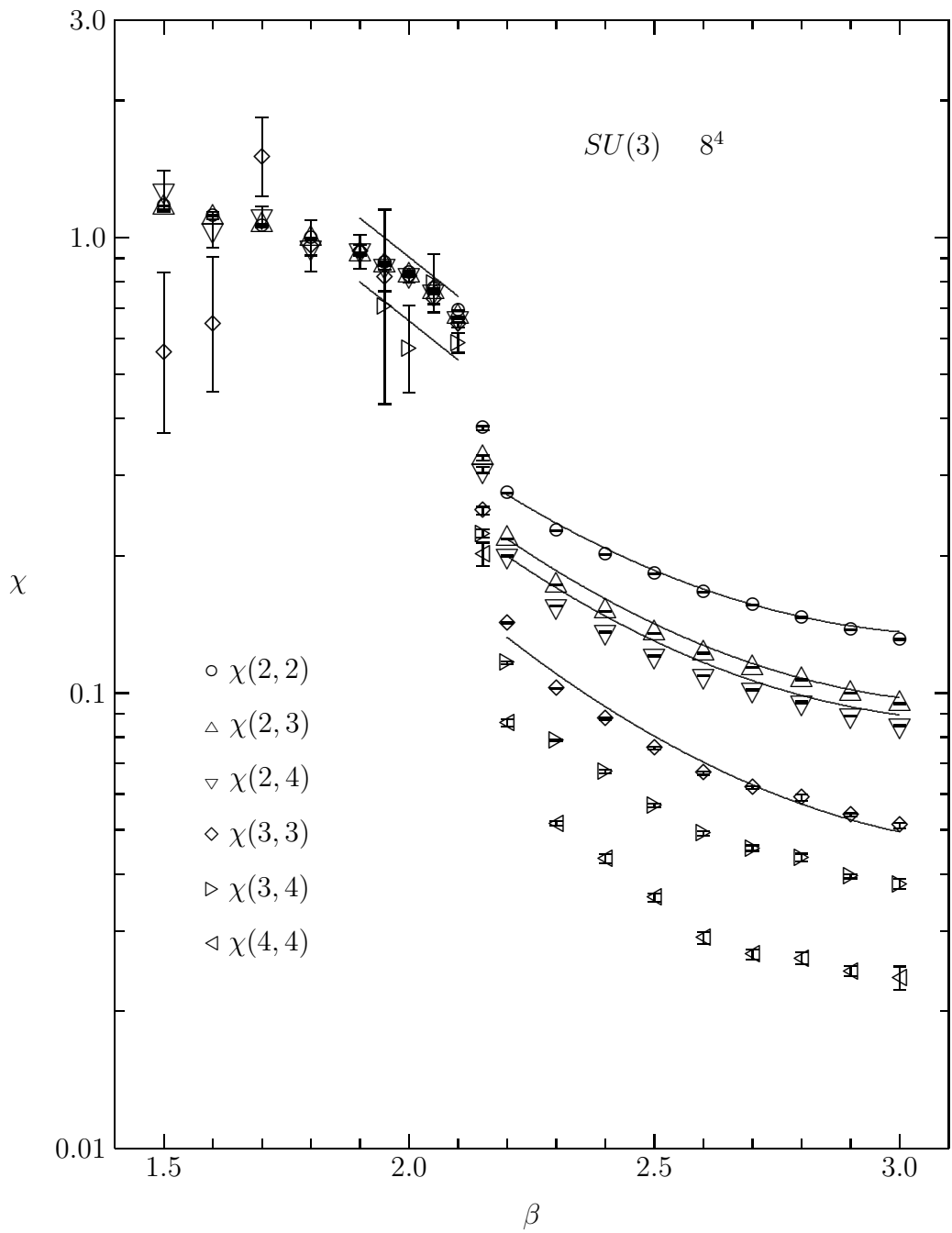
\end{document}